\documentclass[conference]{IEEEtran}
\usepackage{amsmath}
\usepackage{latexsym}
\usepackage{amssymb}
\usepackage{comment}
\usepackage{url}
\usepackage[pdftex]{graphicx}
\usepackage[usenames]{color}
\usepackage{capt-of}
\usepackage{program}
\usepackage{xspace}
\usepackage{sidecap}
\usepackage{subfigure}
\usepackage{booktabs}

\newcommand{\solver}{\textsc{cvc}{\small 4}\textsc{sy}\xspace}

\usepackage[
hidelinks,
pdftex,
pdftitle={CVC4SY for SyGuS-COMP 2019},
pdfauthor={Andrew Reynolds, Haniel Barbosa, Andres N\"otzli, Clark Barrett, Cesare Tinelli},
pdfkeywords={},
pdfborder={0 0 0},
draft=false,
bookmarksnumbered,
bookmarks,
bookmarksdepth=2,
bookmarksopenlevel=2,
bookmarksopen]{hyperref}

\newcommand{\cvc}{\textsc{cvc}{\small 4}\xspace}

\newcommand{\define}[1]{\textsl{#1}}

\newcommand{\con}[1]{\mathsf{#1}}

\newcommand{\Int}{\con{Int}}
\newcommand{\ite}{\con{ite}}
\newcommand{\sconcat}{\con{concat}}

\usepackage{cite}

\def\negvthinspace{\kern-0.083333em}
\def\vthinspace{\kern+0.083333em}
\def\vvthinspace{\kern+0.0416667em}
\def\negvvthinspace{\kern-0.0416667em}

\begin{document}
\pagenumbering{gobble}
\pagestyle{plain}

\title{\textsc{cvc}4\textsc{sy} for SyGuS-COMP 2019}

\author{\IEEEauthorblockN{Andrew Reynolds, Haniel Barbosa, Cesare Tinelli}
\IEEEauthorblockA{The University of Iowa}
\and
\IEEEauthorblockN{Andres N\"otzli, Clark Barrett}
\IEEEauthorblockA{Stanford University}
}

\date{}

\maketitle

\begin{abstract}
  \solver is a syntax-guided synthesis (SyGuS) solver based on bounded term
  enumeration and, for restricted fragments, quantifier elimination.
  The enumerative strategies are based on encoding term enumeration as an
  extension of the quantifier-free theory of algebraic datatypes and on a highly
  optimized brute-force algorithm.
  The quantifier elimination strategy extracts solutions from unsatisfiability
  proofs of the negated form of synthesis conjectures. It uses recent
  counterexample-guided techniques for quantifier instantiation that make
  finding such proofs practically feasible.
  \solver implements these strategies by extending the satisfiability modulo
  theories (SMT) solver \cvc. The strategy to be applied on a given problem is
  chosen heuristically based on the problem's structure.
  This document gives an overview of these techniques and their implementation
  in the SyGuS Solver \solver, an entry for SyGuS-Comp 2019.\footnote{This work
    was partially supported by the National Science Foundation under award
    1656926 and by the Defense Advanced Research Projects Agency (award
    FA8650-18-2-7854).}
\end{abstract}

\section{A Refutation-Based Approach for Synthesis}

We consider the synthesis problem in the context of some theory $T$.  In this
setting, a \define{synthesis conjecture} is expressed as a formula of the form
\begin{eqnarray} \label{eqn:syn_conj}
\exists f\vthinspace
\forall x_1\, \cdots\, \forall x_n.\,P[f, x_1, \ldots, x_n]
\end{eqnarray}
where the second-order variable $f$ represents
the function to be synthesized and $P$ is a formula encoding
properties that $f$ must satisfy for all possible values of the input tuple
$\bar x = (x_1,\ldots,x_n)$.
In this setting, finding a witness for this satisfiability problem amounts
to finding a function for $f$ in some model of $T$
that satisfies $\forall \bar x.\, P[f, \bar x]$.

To determine the satisfiability
of $\exists f\vthinspace \forall \bar x.\, P[f, \bar x]$
an SMT solver~\cite{Barrett2009} can consider the satisfiability
of the (open) formula $\forall \bar x.\, P[f, \bar x]$ by treating $f$
as an uninterpreted function symbol.
We instead use an approach to synthesis geared toward establishing
the \emph{unsatisfiability of its negation}:
\begin{eqnarray} \label{eqn:neg_syn_conj}
  \forall\negvthinspace f\vthinspace\exists \bar x.\, \lnot P[f, \bar x]
\end{eqnarray}
%
A syntactic solution for (\ref{eqn:syn_conj}) can be constructed from a refutation
of (\ref{eqn:neg_syn_conj}),
as opposed to being extracted from the valuation of $f$ in a model of $\forall \bar x.\, P[f, \bar x]$.
Proving (\ref{eqn:neg_syn_conj}) unsatisfiable poses a challenge to current SMT solvers, namely,
dealing with the second-order universal quantification of $f$.
We use two specialized methods (Sections~\ref{sec:sg} and~\ref{sec:si}) to refute
negated synthesis conjectures like (\ref{eqn:neg_syn_conj})
that build on existing capabilities of these solvers.
These are described in detail in~\cite{ReynoldsDKBT15Cav,Reynolds2017}.

\subsection{Syntax-Guided Enumeration}
\label{sec:sg}

The first method is general, and follows the \emph{syntax-guided synthesis}
paradigm~\cite{AlurETAL13SyntaxguidedSynthesis} where the synthesis conjecture
is accompanied by an explicit syntactic restriction on the space of possible
solutions.  Similarly to other SyGuS solvers, the syntax-guided search in
\solver~\cite{Reynolds2019} is based on counterexample-guided inductive
synthesis (CEGIS)~\cite{SolarLezamaETAL06CombinatorialSketchingFinitePrograms}:
a refinement loop in which a learner proposes solutions, and a verifier,
generally a satisfiability modulo theories (SMT) solver, checks them and
provides counterexamples for failures.
Generally, the learner enumerates some set of terms, while pruning spurious
ones~\cite{Udupa2013}.
%


Our syntax-guided synthesis method is based on encoding the syntax
of terms as first-order values.  We use a deep embedding into an extension of
the background theory $T$ with a theory of algebraic data types, encoding the
restrictions of a syntax-guided synthesis problem.  In detail, a set of
syntactic restrictions for a function $f$ may be expressed as an algebraic
datatype whose constructors represent the possible operators for $f$.  For
instance, consider the case where $f$ has type $\Int \rightarrow \Int$.  The
algebraic datatype:
\[
 \begin{array}{l@{\quad}l@{\quad}l}
  \con S & := & \con x_1 \mid \con{zero} \mid \con{one} \mid \con{plus}(\con S, \con S)
 \end{array}
\]
encodes a term signature for $f$ that includes nullary
constructors for the variable $x_1$ (the first input argument of $f$),
and constructors for the symbols of the arithmetic theory $T$.
Terms of sort $\con S$ refer to terms of sort $\Int$.
We extend the background theory with an \emph{evaluation operator} $e_{\con S}$
or a function of type $\con S \rightarrow ( \Int \rightarrow \Int )$,
whose semantics evaluates the analog of its first argument on the second argument.
For instance, $e_{\con S}(\con{plus}(\con x_1,\con{one}), 3 ) = 4$.
Now, consider the (negated) synthesis conjecture $\forall\negvthinspace f\vthinspace\exists x.\, \lnot P[f, x]$,
where solutions for $f$ are restricted to the signature described by datatype $\con S$.
This conjecture may be phrased as the (first-order) formula:
\begin{equation} \label{eqn:syn_conj_sg}
\forall z\vthinspace\exists \bar x.\, \lnot P[e_{\con S}(z), \bar x]
\end{equation}
where $z$ has type $\con S$.
The solver may show (\ref{eqn:syn_conj_sg}) unsatisfiable by finding a (single)
term $t$ of type $\con S$ for which $\exists \bar x.\, \lnot P[e_{\con S}(t),
\bar x]$ is unsatisfiable in an extension of $T$.
When this is the case, the solution for $\forall\negvthinspace
f\vthinspace\exists \bar x.\, \lnot P[f, \bar x]$ is constructed by traversing
the structure of $t$, while replacing constructor symbols with their
corresponding symbols from the signature of $T$.

The enumeration performed by \solver is parameterized by an enumeration strategy
chosen before solving:
%
it either applies a constraint-based (\emph{smart}) enumeration, which allows
for numerous optimizations~\cite[Section 2]{Reynolds2019}; a new approach for
(\emph{fast}) enumerative synthesis~\cite[Section 3]{Reynolds2019}, which
applies a highly optimized brute-force algorithm and has significant advantages
with respect to the smart enumeration; and a \emph{hybrid} approach combining
smart and fast enumeration~\cite[Section 4]{Reynolds2019}.
All enumeration strategies rely on techniques that ensure subterms in candidate
solutions are unique up to theory-specific rewriting.  For example, since terms
$x_1$ and $x_1 + 0$ are equivalent, a \emph{symmetry breaking clause} such as
$\neg \con{isplus}( t ) \vee \neg \con{iszero}( t.2 )$ can be used to avoid
enumerating solutions where the second child of $t$, denoted $t.2$, is zero.

\subsection{Quantifier Instantiation for Single-Invocation Properties}
\label{sec:si}

The second method applies to a restricted, but fairly common, case of synthesis conjectures.
When axiomatizing properties of a desired function $f$,
a particularly well-behaved class are \emph{single-invocation properties}.
Similar classes of conjectures have been studied in recent work~\cite{Alur2015,DBLP:conf/tacas/NeiderSM16}.

A \define{single-invocation property} is any formula of the form
$Q[\bar x, f(\bar x)]$ obtained as an instance of a quantifier-free formula
$Q[\bar x, y]$ not containing $f$.
In other words, the only occurrences of $f$ in $Q[\bar x, f(\bar x)]$ are
in subterms of the form $f(\bar x)$
with the \emph{same} tuple $\bar x$ of \emph{pairwise distinct} variables.
The conjecture
$\forall\negvthinspace f\vthinspace\exists \bar x.\, \lnot Q[\bar x, f(\bar x)]$ is
equivalent to the \emph{first-order} formula:
\begin{equation} \label{eqn:syn_conj_no_syntax}
  \exists \bar x\vthinspace\vthinspace \forall y.\, \lnot Q[\bar x, y]
\end{equation}
To prove the unsatisfiability of (\ref{eqn:syn_conj_no_syntax}),
it suffices to find a $T$-unsatisfiable finite set $\Gamma$ of ground instances of $\lnot Q[\bar{\con{k}}, y]$,
where $\bar{\con{k}}$ is a set of fresh uninterpreted constants.
To find such a $\Gamma$,
we use a specialized new technique, which we refer to as \emph{counterexample-guided quantifier instantiation}.
This technique is similar to CEGIS, but with the difference of being built
directly into the SMT solver.
The idea is to choose instantiations for $\forall y.\, \lnot Q[\bar{\con{k}}, y]$ based on models for
$Q[\bar{\con{k}}, \con{e}]$, where $\con{e}$ is a fresh constant, using a \emph{selection function}.
Selection functions are specific to the background theory.
Recent work has shown selection functions for linear real and integer arithmetic~\cite{DBLP:journals/fmsd/ReynoldsKK17}
and implemented them in \solver.
As a consequence of this work, \solver is a complete synthesis procedure for
single-invocation synthesis conjectures over linear arithmetic.

Assuming the solver finds such a $T$-unsatisfiable $\Gamma$, say $\{\lnot Q[\bar{\con k}, t_1[\bar{\con k}]], \ldots, \lnot Q[\bar{\con k}, t_p[\bar{\con k}]]\}$,
it can be shown that:
\begin{equation} \label{eqn:solution}
\lambda \bar x.\, \ite( Q[\bar x, t_p], t_p, (\,\cdots\, \ite( Q[\bar x, t_2], t_2, t_1 ) \,\cdots\, ))
\end{equation}
is a solution for the synthesis conjecture
$\forall f\, \exists \bar x\, Q[\bar x, f(\bar x)]$.
In the case that $f$ has syntactic restrictions,
we use heuristic enumerative techniques to find a function equivalent to (\ref{eqn:solution})
that meets such syntactic restrictions.

\section{Enhancements}
\label{sec:enh}

\subsection{Optimizations for Enumerative Search in SMT}

Since last year, we have added a number of optimizations to \solver's
syntax-guided enumerative search to make it faster~\cite{Reynolds2019}.
A key element of the \emph{smart} enumerative
search in \solver is the use of a theory of datatypes with \emph{shared
  selectors}~\cite{Reynolds2018-shared}.  At a high level, this ensures a
maximal number of (e.g., symmetry breaking) clauses are shared across multiple
contexts of the search.
The new \emph{fast} enumerative search leads to gains in most problem
categories, but is especially impactful on PBE problems, where it outperforms
the smart strategy by several orders of magnitude.
Such gains are significant given that \solver won this track at SyGuS-COMP~2018
by employing the smart technique alone.
We have also made significant improvements to \solver's theory
rewriting~\cite{Notzli2019,Reynolds2019-strings}, which is used to infer when a
term is equivalent to a previous one and hence can be discarded. This benefits
both smart and fast enumerations.


\subsection{Strategy-Based Solution Construction}

\solver uses divide-and-conquer techniques inspired
by~\cite{Alur2015,alur2017scaling} to construct solutions for synthesis
conjectures.
At a high level, the approach enumerates a stream of terms and independently
devises a strategy for combining them into candidate solutions.
When applicable, \solver uses strategies for building solutions involving
$\ite$-terms (to create decision trees), string concatenation $\sconcat$ in the
PBE strings division (to sequence solutions), and combinations of the two.

\subsection{Invariant Synthesis}

For the invariant synthesis track, \solver uses several static preprocessing
passes to make the conjecture easier to solve.
Most importantly, this includes rephrasing the invariant synthesis problem as
finding a \emph{strengthening of the post-condition} (resp.  weakening of the
pre-condition).
In other words, given an invariant synthesis problem for predicate $I$,
instead of synthesizing $I( \bar x )$, \solver may choose to instead
synthesize a predicate $I'( \bar x )$ such that
$post( \bar x ) \wedge I'( \bar x )$ is the overall solution,
where $post( \bar x )$ is the post-condition of the invariant synthesis problem.
%
Additionally, \solver uses a new divide-and-conquer approach, Unif+PI, for function
synthesis that works especially well for invariant
synthesis~\cite{Barbosa2019-unifpi}.
It accumulates refinement lemmas, synthesizes partial solutions for each point
in these lemmas independently, and uses a decision tree algorithm to combine
these partial solutions into an overall solution.

\subsection{Constant Repair}

Since last year, \solver implements new techniques~\cite{Abate2018} for using its theory solvers to
synthesize constants to repair candidate solutions.
To do so, occurrences of $(\con{Constant}\ T)$ in grammars are
treated as a constructor $\con{const}$ containing a single field of type $T$.
If a candidate solution is enumerated that involves this constructor,
a subcall to \cvc is made to find the appropriate constant, if one exists,
such that the candidate solution (universally) satisfies the overall conjecture.
For example, in terms of the deep embedding, if our candidate solution is
$\con{plus}(\con{x},\con{const}(1))$ and our specification is $\exists f\vthinspace
\forall x.\, f(x)>x+100$, then we check the satisfiability of the quantified
conjecture $\exists c\vthinspace\forall x.\, x+c>x+100$, which is satisfiable with a
model where, e.g., $c=101$.

\subsection{Inferring Equivalent Single-Invocation Properties}

CVC4 uses techniques for recognizing when synthesis conjectures can be rewritten into a form that is single invocation.
This includes normalizing the arguments of invocations across conjunctions, and applying quantifier elimination to variables for which
the function to synthesize is not applied.

\subsection{Supporting New SyGuS Language}

This year \solver has been extended to support the new SyGuS language\footnote{https://sygus.org/assets/pdf/SyGuS-IF\_2.0.pdf}, which is a
lightly-modified version of the SyGuS input format that was used in previous
competitions. The new format is more compliant with SMT-LIB version 2.6,
includes minor changes to the concrete syntax for commands, and eliminates
several deprecated features of the previous format.

\section{Configurations for SyGuS-COMP 2019}

\newcommand{\strat}[1]{{\bf #1}}

\solver is entering all tracks of SyGuS-COMP 2019.  For all tracks, \solver
performs the following steps, given a background theory $T$:

\begin{enumerate}
\item Parse the input and phrase the problem in terms of a (negated) synthesis conjecture of the form $\forall\negvthinspace f\vthinspace\exists \bar x.\, \lnot P[f, \bar x]$,
      and a set of syntactic restrictions $R$.
\item Do one of the following:
\begin{enumerate}
\item Determine a single-invocation property $Q[ \bar y, f( \bar y ) ]$ that is equivalent to $P[f, \bar x]$.
      Let $\varphi$ be $\forall \bar z.\, Q[ \bar{\con{k}}, \bar z]$ and let $T'$ be $T$.
\item Let $\varphi$ be $\forall z\vthinspace\exists \bar x.\, P[ e_{\con S}(z), \bar x]$,
      where $\con S$ is an algebraic datatype that encodes $R$, 
      and let $T'$ be an extension of $T$ whose signature includes $\con S$ and $e_{\con S}$, as described in Section~\ref{sec:sg}.
\end{enumerate}
\item Using the appropriate technique (either the one from Section~\ref{sec:sg} or~\ref{sec:si}), show that $\varphi$ is $T'$-unsatisfiable.
\item If successful, reconstruct a solution for the original synthesis conjecture $\forall\negvthinspace f\vthinspace\exists \bar x.\, \lnot P[f, \bar x]$.
\end{enumerate}

For the GENERAL and CLIA tracks, \solver prefers executing Step 2a over Step
2b.  In the case that \solver executes Step 2a but fails to reconstruct a
solution in Step 4, \solver restarts and executes Step 2b instead.


For some tracks, we enter different configurations with different synthesis
strategies.
We denote the \emph{smart} enumerative strategy by \strat{s} and the
\emph{fast} enumerative strategy by \strat{f}.
The Unif+PI strategy for invariant synthesis is denoted by
\strat{su}. We denote the the auto strategy, which heuristically picks a
strategy based on the properties of the problem, by \strat{auto}. It uses the
single-invocation solver on problems that are amenable to quantifier
elimination, strategy \strat{f} on PBE problems and problems without the Boolean
type or the $\ite$ operator in their grammar and strategy \strat{s} otherwise.

\solver has the following entries, per track, in SyGuS-COMP 2019:
\begin{center}
  \begin{tabular}{ll}
    \toprule
    Track&Configuration\\
    \midrule
    CLIA& \strat{auto}\\
    GENERAL& \strat{auto}, \strat{f}, \strat{s}\\
    INV& \strat{f}, \strat{s}, \strat{su}\\
    PBE\_BitVec& \strat{f}, \strat{s}\\
    PBE\_Strings& \strat{f}, \strat{s}\\
    \bottomrule
  \end{tabular}
\end{center}

\bibliographystyle{abbrv}
\bibliography{cvc4}

\begin{thebibliography}{10}

\bibitem{Abate2018}
A.~Abate, C.~David, P.~Kesseli, D.~Kroening, and E.~Polgreen.
\newblock Counterexample guided inductive synthesis modulo theories.
\newblock In H.~Chockler and G.~Weissenbacher, editors, {\em Computer Aided
  Verification (CAV), Part I}, volume 10981 of {\em Lecture Notes in Computer
  Science}, pages 270--288. Springer, 2018.

\bibitem{AlurETAL13SyntaxguidedSynthesis}
R.~Alur, R.~Bod{\'{\i}}k, G.~Juniwal, M.~M.~K. Martin, M.~Raghothaman, S.~A.
  Seshia, R.~Singh, A.~Solar{-}Lezama, E.~Torlak, and A.~Udupa.
\newblock Syntax-guided synthesis.
\newblock In {\em FMCAD}, pages 1--17. {IEEE}, 2013.

\bibitem{Alur2015}
R.~Alur, P.~{\v{C}}ern{\'y}, and A.~Radhakrishna.
\newblock {\em Synthesis Through Unification}, pages 163--179.
\newblock Springer International Publishing, Cham, 2015.

\bibitem{alur2017scaling}
R.~Alur, A.~Radhakrishna, and A.~Udupa.
\newblock Scaling enumerative program synthesis via divide and conquer.
\newblock In {\em International Conference on Tools and Algorithms for the
  Construction and Analysis of Systems}, pages 319--336. Springer, Berlin,
  Heidelberg, 2017.

\bibitem{Barbosa2019-unifpi}
H.~Barbosa, A.~Reynolds, D.~Larraz, and C.~Tinelli.
\newblock Extending enumerative function synthesis via~{SMT}-driven
  classification.
\newblock In C.~Barrett and J.~Yang, editors, {\em Formal Methods In
  Computer-Aided Design (FMCAD)}. (Accepted for publication). {IEEE}, 2019.

\bibitem{Barrett2009}
C.~Barrett, R.~Sebastiani, S.~Seshia, and C.~Tinelli.
\newblock {Satisfiability Modulo Theories}.
\newblock In A.~Biere, M.~J.~H. Heule, H.~van Maaren, and T.~Walsh, editors,
  {\em Handbook of Satisfiability}, volume 185 of {\em Frontiers in Artificial
  Intelligence and Applications}, chapter~26, pages 825--885. IOS Press, 2009.

\bibitem{DBLP:conf/tacas/NeiderSM16}
D.~Neider, S.~Saha, and P.~Madhusudan.
\newblock Synthesizing piece-wise functions by learning classifiers.
\newblock In M.~Chechik and J.~Raskin, editors, {\em Tools and Algorithms for
  the Construction and Analysis of Systems - 22nd International Conference,
  {TACAS} 2016, Held as Part of the European Joint Conferences on Theory and
  Practice of Software, {ETAPS} 2016, Eindhoven, The Netherlands, April 2-8,
  2016, Proceedings}, volume 9636 of {\em Lecture Notes in Computer Science},
  pages 186--203. Springer, 2016.

\bibitem{Notzli2019}
A.~N\"otzli, A.~Reynolds, H.~Barbosa, A.~Niemetz, M.~Preiner, C.~Barrett, and
  C.~Tinelli.
\newblock Syntax-guided rewrite rule enumeration for {SMT} solvers.
\newblock In M.~Janota and I.~Lynce, editors, {\em Theory and Applications of
  Satisfiability Testing (SAT)}, (Accepted for publication). Lecture Notes in
  Computer Science. Springer, 2019.

\bibitem{Reynolds2019}
A.~Reynolds, H.~Barbosa, A.~N{\"o}tzli, C.~Barrett, and C.~Tinelli.
\newblock cvc4sy: Smart and fast term enumeration for syntax-guided synthesis.
\newblock In I.~Dillig and S.~Tasiran, editors, {\em Computer Aided
  Verification (CAV), Part II}, volume 11562 of {\em Lecture Notes in Computer
  Science}, pages 74--83, Cham, 2019. Springer International Publishing.

\bibitem{ReynoldsDKBT15Cav}
A.~Reynolds, M.~Deters, V.~Kuncak, C.~W. Barrett, and C.~Tinelli.
\newblock Counterexample guided quantifier instantiation for synthesis in
  {CVC4}.
\newblock In {\em Computer Aided Verification (CAV)}. Springer, 2015.

\bibitem{DBLP:journals/fmsd/ReynoldsKK17}
A.~Reynolds, T.~King, and V.~Kuncak.
\newblock Solving quantified linear arithmetic by counterexample-guided
  instantiation.
\newblock {\em Formal Methods in System Design}, 51(3):500--532, 2017.

\bibitem{Reynolds2017}
A.~Reynolds, V.~Kuncak, C.~Tinelli, C.~Barrett, and M.~Deters.
\newblock Refutation-based synthesis in smt.
\newblock {\em Formal Methods in System Design}, 2017.

\bibitem{Reynolds2019-strings}
A.~Reynolds, A.~N{\"{o}}tzli, C.~W. Barrett, and C.~Tinelli.
\newblock High-level abstractions for simplifying extended string constraints
  in {SMT}.
\newblock In I.~Dillig and S.~Tasiran, editors, {\em Computer Aided
  Verification (CAV), Part II}, volume 11562 of {\em Lecture Notes in Computer
  Science}, pages 23--42. Springer, 2019.

\bibitem{Reynolds2018-shared}
A.~Reynolds, A.~Viswanathan, H.~Barbosa, C.~Tinelli, and C.~Barrett.
\newblock Datatypes with shared selectors.
\newblock In D.~Galmiche, S.~Schulz, and R.~Sebastiani, editors, {\em
  International Joint Conference on Automated Reasoning (IJCAR)}, volume 10900
  of {\em Lecture Notes in Computer Science}, pages 591--608. Springer, 2018.

\bibitem{SolarLezamaETAL06CombinatorialSketchingFinitePrograms}
A.~Solar{-}Lezama, L.~Tancau, R.~Bod{\'{\i}}k, S.~A. Seshia, and V.~A.
  Saraswat.
\newblock Combinatorial sketching for finite programs.
\newblock In J.~P. Shen and M.~Martonosi, editors, {\em ASPLOS}, pages
  404--415. {ACM}, 2006.

\bibitem{Udupa2013}
A.~Udupa, A.~Raghavan, J.~V. Deshmukh, S.~Mador-Haim, M.~M. Martin, and
  R.~Alur.
\newblock Transit: Specifying protocols with concolic snippets.
\newblock In {\em PLDI}, pages 287--296. ACM, 2013.

\end{thebibliography}

\end{document}